\documentclass[12pt,preprint]{aastex}

\slugcomment{Version \today}

\shorttitle{Oblique Ion Two-Stream Instability}
\shortauthors{Ohira and Takahara}

\begin{document}

\title{Oblique Ion Two-Stream Instability in the Foot Region of 
a Collisionless Shock}

\author{Yutaka Ohira and Fumio Takahara}
\affil{Department of Earth and Space Science, Graduate School
of Science, Osaka University, Machikaneyama 1-1, Toyonaka, 
Osaka 560-0043, Japan \\
{\rm yutaka@vega.ess.sci.osaka-u.ac.jp,
 takahara@vega.ess.sci.osaka-u.ac.jp}}

\begin{abstract}
Electrostatic behavior of a collisionless plasma in the foot region 
of high Mach number perpendicular shocks is investigated through 
the two-dimensional linear analysis and electrostatic 
particle-in-cell (PIC) simulation. The simulations are double 
periodic and taken as a proxy for the situation in the foot. 
The linear analysis for relatively cold unmagnetized plasmas with 
a reflected proton beam shows that obliquely propagating Buneman 
instability is strongly excited. We also found that when the electron 
temperature is much higher than the proton temperature, the most 
unstable mode is the highly obliquely propagating ion two-stream 
instability excited through the resonance between ion plasma 
oscillations of the background protons and of the beam protons, 
rather than the ion acoustic instability that is dominant for 
parallel propagation. 
 
To investigate nonlinear behavior of the ion two-stream instability, 
we have made PIC simulations for the shock foot region in which the 
initial state satisfies the Buneman instability condition. 
In the first phase, electrostatic waves grow two-dimensionally by the 
Buneman instability to heat electrons. In the second phase, 
highly oblique ion two-stream instability grows to heat mainly ions. 
This result is in contrast to previous studies based on one-dimensional 
simulations, for which ion acoustic instability further heats electrons.  
 
The present result implies that overheating problem of electrons for 
shocks in supernova remnants is resolved by considering ion two-stream 
instability propagating highly obliquely to the shock normal and that 
multi-dimensional analysis is crucial to understand the particle heating 
and acceleration processes in shocks.
\end{abstract}

\keywords{supernova remnants -- shock waves -- plasmas -- instabililes --
cosmic rays -- acceleration of particle}

\section{Introduction}
The discovery of thermal and synchrotron X-rays from young supernova 
remnants (SNRs) provides the evidence that electrons are heated up to 
a few keV and that a portion of them are accelerated to highly 
relativistic energy in SNR shocks \citep{koy95}. Because SNR shocks 
are collisionless, not only particle acceleration mechanisms but also 
electron heating mechanisms in SNR shocks are not so simple. Previous 
studies have given an important key to the formation mechanism of 
perpendicular collisionless shocks. When the Alfv${\rm \acute{e}}$n Mach 
number $M_{\rm A}$ is larger than the critical Mach number, about 3, a 
perpendicular shock reflects some of the incident ions to the upstream, 
where a foot region forms on a spatial scale of the ion gyroradius 
\citep{ler83}. The plasma in the foot region consists of incident ions 
and electrons and reflected ions and returning ions which are made from 
reflected ions and move to the shock after a gyration. As for the 
electron heating machanism, \citet{pap88} proposed that when the Mach 
number is larger than $0.5(m_{\rm p}/m_{\rm e})^{1/2}\sim 20$, incident 
electrons and reflected ions excite electrostatic waves by the Buneman 
instability \citep{bun58} because the relative velocity between them is 
large compared with the electron thermal velocity. They also suggest 
that after electrons are  heated by electrostatic waves induced by the 
Buneman instability, ion acoustic instability is triggered because the 
electron temperature becomes much higher than the proton temperature. 
As a result, electrons are strongly heated by the Buneman instability 
and the ion acoustic instability. \citet{car88} performed a 
one-dimensional hybrid simulation and demonstrated that strong electron 
heating actually occurs. They concluded that an $M_{\rm A}=500$ shock 
heats electrons by a factor of $10^5$ across the shock. This means that 
if the upstream electron temperature is 1 eV, the downstream electron 
temperature becomes 100 keV. This value of the downstream temperature 
is much larger than the recent observational one for SNRs, a few keV 
\citep{sta06}. This discrepancy has been an open issue to be resolved 
for a long time.

On the other hand, \citet{shi00} and \cite{hos02} performed 
one-dimensional full particle-in-cell (PIC) simulations to investigate 
the electron acceleration at perpendicular shocks. Their simulation 
solves a whole region of a collisionless perpendicular shock and makes 
reflected ions self-consistently by employing a small proton to 
electron mass ratio. Their results showed that electrons are not only 
heated at the foot region but also significantly accelerated by surfing 
acceleration mechanism. However, this acceleration is valid only for 
the one-dimensional case because the surfing acceleration strongly 
depends on the structure of the electrostatic potential. In our first 
paper \citep{ohi07}, we performed two-dimensional electrostatic PIC 
simulations to solve for the two-dimensinal structure of the 
electrostatic potential excited by the Buneman instability. We 
employed the real mass ratio but the simulation region is limited to 
the foot region. Our results showed that oblique modes grow as strongly 
as the modes parallel to the beam direction, that the potential 
structure becomes two-dimensional and that no efficient surfing 
acceleration occurs, while electron heating occurs. Thus, the problem 
of electron acceleration has been back to the start again. In that 
paper, we concentrated on the stage of the Buneman instability and did 
not follow the long time scale evolution after the Buneman nstability 
has saturated.

In this paper, we study the time evolution of electrostatic collisionless 
plasma instabilities in the foot region by making linear analysis and 
by performing two-dimensional electrostatic PIC simulation. We perform 
simulations with a higher resolution, a larger simulation box and a 
longer simulation time than in \citet{ohi07}. Especially, we focus on 
the evolution after electrostatic waves excited by the Buneman 
instability have decayed. Our simulation substantially improves most 
previous works that are one-dimensional, employ an artifically small 
proton electron mass ratio or impose rather strong magnetic field. It 
is obvious that a multi-dimensional analysis is necessary as discussed 
above \citep{blu60, lam74, ohi07}. Our motivation for employing the real 
mass ratio is as follows. For a small mass ratio, the foot region in 
the simulation is shorter than the realistic one and the time scale on 
which electrons stay in the foot region in the simulation is also 
shorter than the realistic one because the size of the foot region is 
about the ion gyroradius $m_{\rm p}v_{\rm d}c/eB$, where $v_{\rm d}$ 
and $B$ are the drift velocity of reflected protons and the magnetic 
field, respectively. Because reflected ions have a large free energy, 
we expect that more energy is transported to electrons through 
collective instabilities with the realistic mass ratio in the foot region. 
The drift velocity is not large enough to excite electromagnetic 
waves, so that electrostatic waves are more important.

In \S 2 we perform linear analysis for two-dimensional electrostatic 
modes. In \S 3 we describe the initial setting of the PIC simulations 
and numerical results, followed by a discussion in \S 4.

\section{Linear analysis}
In this section, we perform linear analysis for two-dimensional 
electrostatic modes in unmagnetized plasmas with beams. In the foot 
region of perpendicular shocks of SNRs, we regard that there are several 
beams with a finite temperature and that their relative velocities are 
much smaller than the light speed. Therefore, the fastest growing modes 
are electrostatic modes. Then, we here concentrate on the electrostatic 
modes. For typical interstellar medium, the ratio 
\begin{equation}
\frac{ \Omega_{\rm ce} }{ \omega_{\rm pe} } 
\simeq 10^{-3} \left(\frac{B}{3\rm{\mu G}}\right) 
\left(\frac{n_{\rm e}}{1\rm{cm}^{-3}}\right)^{-1/2}
\end{equation}
is relatively small, where $\Omega_{\rm ce}, \omega_{\rm pe}$ and 
$n_{\rm e}$ are electron cyclotron frequency, electron plasma frequency, 
and electron number density, respectively. So plasma oscillations are 
hardly changed by the magnetic field in the foot region of shocks of 
SNRs. When we consider spatial scale smaller than the gyroradius, 
we may neglect the effects of magnetic fields. Thus, we concentrate here 
on unmagnetized plasmas. 

We define such that the $x$-direction is shock normal direction and 
the $y$-direction is the direction that is perpendicular to shock normal 
and wave vectors are on the $x-y$ plane. For unmagnetized collisionless 
plasmas, the electrostatic dispersion relation reads as
\begin{eqnarray}
&&1+\sum_{\rm s} \frac{\omega_{\rm ps}^2}{k^2}
\int d^2 v \frac{ {\bf k} \cdot {\bf \nabla}_v f_{\rm s0}}
{\omega - {\bf k} \cdot {\bf v}} = 0,  \\
&&k=\sqrt{k_{x}^2+k_{y}^2},
\end{eqnarray} 
where the subscript s represents particle species, here electrons, ions 
and beam ions, $\omega_{\rm ps}=(4\pi n_{\rm s}e^2/m_{\rm s})^{1/2}$ is 
the plasma frequency of the particle species s and $f_{\rm s0}$ is the 
normalized distribution function of the particle species s,
\begin{equation} 
f_{\rm s0}
=\frac{1}{\pi v_{\rm th,s}^2} 
\exp \left[-\frac{(v_{x}-v_{\rm d,s})^2+v_{y}^2}
{v_{\rm th,s}^2}\right]
\end{equation} 
where $v_{\rm th,s}=(2kT_{\rm s}/m_{\rm s})^{1/2}$ and $v_{\rm d,s}$ is 
the thermal velocity and drift velocity of the particle species s, 
respectively. 

To make equation (2) simpler, we use new coordinates $x'$ and $y'$ as 
\begin{eqnarray}
&& x'=x \cos \theta+y \sin \theta , \nonumber \\
&& y'=-x \sin \theta +y \cos \theta ,\\
&& \cos \theta = \frac{k_{x}}{\sqrt{k_{x}^2 + k_{y}^2}}. 
\nonumber
\end{eqnarray}
Then, equations (2) and (3) become 
\begin{equation}
1+\sum_{\rm s} \frac{\omega_{\rm ps}^2}{k}
\int d^2 v' \frac{\partial f_{\rm s0}/\partial v'_{x}}
{\omega - k v'_{x}} = 0,
\end{equation}
and
\begin{equation}
f_{\rm s0}=\frac{1}{\pi v_{\rm th,s}^2} 
\exp \left[-\frac{(v'_{x}-v_{\rm d,s}\cos \theta)^2
+(v'_{y}+v_{\rm d,s}\sin \theta)^2}{v_{\rm th,s}^2}\right].
\end{equation}
Finally, we substitute equation (7) into  equation (6), and we obtain
\begin{eqnarray}
&&1+\sum_{\rm s} \frac{2\omega_{\rm ps}^2}{k^2v_{\rm th,s}^2}
\left[1+\xi_{\rm s} Z(\xi_{\rm s})\right]=0, \nonumber \\
&&Z(\xi_{\rm s}) = \frac{1}{\sqrt{\pi}}
\int_{-\infty}^{\infty}\frac{e^{-z^2}}{z-\xi_{\rm s}}dz, \\
&&\xi_{\rm s} = \frac{\omega-k_{x}v_{\rm d,s}}{kv_{\rm th,s}}, \nonumber
\end{eqnarray}
where $Z(\xi_{\rm s})$ is the plasma dispersion function and it can be 
numerically solved \citep{wata91}. 

We here present results of the linear analysis about two cases of plasma 
conditions and discuss on three kinds of plasma instabilities. 

\subsection{Buneman instability}

We first consider the situation in which there are three beams, incident 
protons, incident electrons and reflected protons and the temperatures 
of all plasma beams are low, typically around 1 eV. We thus we neglect 
the contribution of returning ions in the dispersion relation, for 
simplicity. As is easily understood, the returning component plays a role 
in assuring the vanishing net current in the unperturbed state. In the 
dispersion relation, they play a symmetrical role to the reflected ions 
and their effects are starightforwardly understood when we make clear 
the role of reflected ions. We make analyses in the upstream rest frame 
in which only reflected ions have a drift velocity, typically 
$v_{\rm d}=v_{\rm d,ref}=0.02c=2v_{\rm sh}$, where $v_{\rm sh}$ is the 
shock velocity. Hence, a typical velocity ratio is 
$v_{\rm d}/v_{\rm th,e}=10$. We assume that the proton reflection ratio 
is $n_{\rm ref}/n_{\rm p}=0.25$. 

The growth rate obtained by solving the linear dispersion relation is 
displayed in Figure 1(a). In this condition, the most unstable mode is 
the Buneman instability. The Buneman instability is caused by the 
resonance between the electron plasma oscillation of the upstream 
electrons and proton plasma oscillation of the reflected proton beam. 
In Figure 1, $k_{x}$ and $k_{y}$ are wavenumbers normalized by 
$\omega_{\rm pe}/v_{\rm d}$ and the color contours show the growth rate 
normalized by $\omega_{\rm pe}$, where only the growth rate of growing 
modes is shown. As is seen, the growth rate of obliquely propagating 
modes is as large as that of modes parallel to the beam direction.  
This feature of the Buneman instability can be well understood in the 
cold limit. We present the results of the cold limit for which all 
temperatures are set to zero in Figure 1(b). In the cold limit, the 
distribution function (4) becomes 
\begin{equation}
f_{\rm s0}=\delta(v_{x}-v_{\rm d,s})\delta(v_{y}),
\end{equation}
and the dispersion relation (2) is reduced to
\begin{equation}
1-\sum_{\rm s} \left( \frac{\omega_{\rm ps}}
{\omega-k_{x} v_{\rm d,s}} \right)^2 = 0.
\end{equation} 
Because $k_{y}$ does not appear in equation (10), the growth rate of 
electrostatic instabilities in the cold limit does not depend on 
$k_{y}$. Therefore, in the cold limit, excited waves have any $k_{y}$ 
and the structure of electrostatic potential to the $y$-direction is 
strongly disordered and loses coherence to the $y$ direction. This 
feature is very important to negate the electron surfing acceleration 
mechanism \citep{ohi07}. The maximum growth rate of the Buneman 
instability in the cold limit is \citep{bun58}
\begin{equation}
\gamma_{\rm max} 
= \left[\frac{3\sqrt{3}}{16} 
\left(\frac{m_{\rm e}}{m_{\rm p}}\right) 
\left(\frac{n_{\rm ref}}{n_{\rm e}}\right)\right]^{1/3}
\omega_{\rm pe}  \ \ 
{\rm at} \ \ k_x =\frac{\omega_{\rm pe}}{v_{\rm d}}
=k_{\rm Bun}. \nonumber 
\end{equation}

In reality, because of a finite temperature, the modes with large 
wavenumbers are suppressed to grow. This is seen in Figure 1(a); 
the growth rate for large $k_y$ decreases. The dispersion relation of 
the Buneman instability depends on $v_{\rm th,e}/v_{\rm d}$ and the 
number density ratio. When $v_{\rm th,e}/v_{\rm d}$ is small, modes 
that have a large $k_{y}$ can grow as long as the wavelength is larger 
than the  electron Debye length. The wavenumber corresponding to the 
electron Debye length is 
$k_{\rm D,e}= \omega_{\rm pe}/v_{\rm th,e}=10k_{\rm Bun}$ 
in the present plasma condition and the boundary between 
growing and damping region in Figure 1(a) is about 
$k_{y} =7k_{\rm Bun}\sim k_{\rm D,e}$, as is consistent with the present 
consideration. This result implies that in SNRs condition, oblique 
modes propagating to the beam direction, i.e., to the shock normal, 
can grow as strongly as the modes to the parallel direction. 
Because these modes are electrostatic, the direction of the excited 
electric fields is  to the wave vector and the energy density of the 
electric field of the $y$-component can be larger than that of 
the $x$-component. 

\subsection{Ion two-stream instability and ion acoustic instability}
After the upstream electrons are heated by the Buneman instability, 
the thermal velocity of electrons increases up to 
$v_{\rm th,e} \sim v_{\rm d}$ while the thermal velocity of protons 
is roughly the same as the initial one, so that the situation of 
$T_{\rm e} \gg T_{\rm p}$ is realized. The other parameters are the 
same as in the low temperature case described in the previous 
subsection. In this condition, while the Buneman instability is 
stabilized, other types of instabilities can occur. We found that 
in addition to the ion acoustic instability discussed previously 
\citep{car88}, the ion two-stream instability that has not been 
well noticed in the literature becomes unstable. The ion two-stream 
instability is caused by the resonance of the ion plasma oscillation 
of the upstream plasma and that of reflected ions in the situation 
$T_{\rm e} \gg T_{\rm p}$. Thus, it occurs concurrently with the ion 
acoustic instability which is caused by the resonance between ion 
acoustic waves of the reflected protons and electron plasma oscillation 
of the upstream plasma, where the former modes are mediated by the 
presence of the hot upstream electrons. 

The numerical results of the growth rate for ion two-stream instability 
and ion acoustic instability are shown in Figures 2(a) and (b), 
respectively. In Figure 2, $k_{x}$, $k_{y}$ and the growth rate are 
normalized in the same way as in Figure 1, and the color contours show 
the growth rate of the growing modes. Note that the difference of the 
range of the wavenumber in the $x$-direction $k_x$ between (a) and (b). 
While ion acoustic instability grows at around 
$k_x\approx \omega_{\rm pe}/v_{\rm d}$, ion two-stream instabily grows 
at much smaller $k_y$. It is also noted that ion two-stream instability 
is seen for low but finite values of $k_y$, In the present conditions, 
the wavenumber corresponding to the ion Debye length is 
$k_{\rm D,p} = \omega_{\rm pi}/v_{\rm th,i}=10k_{\rm Bun}$ while that to the 
electron Debye length is $k_{\rm Bun}$. Ion plasma oscillation exists 
between these two wavenumbers. For wavenumbers lower than $k_{\rm Bun}$, 
it reduces to the ion acoustic mode while for wavenumbers higher than 
$k_{\rm D,p}$ it damps by thermal motions of ions. It should be noted 
that in Figure 2(a), ion two-stream modes of parallel propagation to the 
beam direction are only weakly growing, but that highly oblique modes grow 
very fast and the maximum growth rate is larger than that of the ion 
acoustic instability by a factor of a few. The reason is explained as 
follows. To excite the ion two-stream instability, the resonance 
condition, $k_{x} v_{\rm d} \sim \omega_{\rm pi}$ must be satisfied in 
addition to the wavenumber condition mentioned above. The resonance 
condition requires a small 
$k_x \approx (m_{\rm e}/m_{\rm p})^{1/2}k_{\rm Bun}$. For propagation 
parallel to the beam direction, this is imcompatible with the wavenumber 
condition and the growth rate is very small. In contrast, when the wave 
has a large $k_{y}$, both the wavenumber condition and the resonance 
condition are fulfilled simultaneously and a larger growth rate is obtained. 
The results shown in Figure 2 (a) are fully consistent with this picture 
of the ion two-stream instability.

To understand the ion two-stream instability through the dispersion 
relation, we consider a situation where the proton temperature 
is zero and the electron temperature is very high and $k_y \gg k_x$ 
(if the electron drift velocity is much smaller than thermal velocity, 
we do not need the final condition). Then, we can approximate as  
$|\xi_{\rm p}|, |\xi_{\rm ref}| \gg 1$ and $|\xi_{\rm e}| \ll 1$, 
and the dispersion relation becomes
\begin{equation}
1+2\left(\frac{k_{\rm D,e}}{k}\right)^2- \left( \frac{\omega_{\rm pp}}
{\omega
} \right)^2-\left( \frac{\omega_{\rm ref}}
{\omega-k_{x} v_{\rm d,ref}} \right)^2= 0.
\end{equation} 
The second term represents the Debye shielding effect of hot electrons 
and becomes small for $k \gg k_{\rm D,e}$ as discussed above. 
It is seen that for $k \gg k_{\rm D,e}$, the dispersion relation has 
the same form as equation (10) and we obtain the maximum growth rate of 
the ion two-stream instability when $n_{\rm ref} \ll n_{\rm p}$, 
by replacing $\omega_{\rm pe}$ with $\omega_{\rm pp}$ in equation (11), as 
\begin{equation}
\gamma_{\rm max} 
= \left[\frac{3\sqrt{3}}{16} 
\left(\frac{n_{\rm ref}}{n_{\rm p}}\right)\right]^{1/3}
\omega_{\rm pp}  \\ 
{\rm at} \ \ k_x =\frac{\omega_{\rm pp}}{v_{\rm d}}. \nonumber 
\end{equation}
Because the upstream plasma stays in foot region by about the proton 
gyro-period $\Omega_{\rm cp}^{-1}$ and because 
$\Omega_{\rm cp}/\omega_{\rm pp}\sim 4.3 \times 10^{-5}$, ion two-stream 
instability can grow enough in the foot region. 

As far as we are aware, this oblique unstable mode has not been 
considered up to now. We expect that this instability heats ions and that 
much affects subsequent electron heating processes. 

\section{Simulation}
To perform two-dimensional simulations with real proton electron mass 
ratio, we confine our attention to the foot region through a proper 
modeling instead of solving the whole shock structure. Our simulation 
box is taken to be at rest in the upstream frame of reference, i.e., 
that of incident protons and electrons. We do not solve electromagnetic 
waves and concentrate on electrostatic waves.

\subsection{Setting}
We define the $x$-direction as the shock normal pointing to the shock 
front, and thus the reflected protons move in the $-x$-direction and 
returning protons move in the $x$-direction. The magnetic field is 
taken to be spatially homogeneous pointing in the $z$-direction and 
we solve the particle motion and electric field in the $x-y$ plane. 
As the initial condition, we prepare upstream electrons, upstream 
protons, reflected protons and returning protons. Each population is 
uniformly distributed in the $x-y$ plane and their momentum 
distribution is given by a Maxwellian at the same temperatures 
$T=T_{\rm e}=T_{\rm p}=T_{\rm ref}=T_{\rm ret}=1.75{\rm eV}$. In 
addition, reflected and returning protons have an extra drift velocity 
in the $x$-direction of $v_{\rm d}=\pm0.04c$ ($v_{\rm d}=2v_{\rm sh}$). 
The number densities of each population are taken as 
$n_{\rm e}=1.5n_{\rm p}=1\rm{cm}^{-3}$ and $n_{\rm ref}
=n_{\rm ret}=0.25n_{\rm p}$, where subscripts e, p, ref and ret 
represent upstream electrons, upstream protons, reflected protons and 
returning protons, respectively (see Figure 3). These parameters are 
typical of young SNRs and satisfy the charge neutrality and a vanishing 
current. 
 
We employ the periodic boundary condition both in the $x$- and 
$y$-directions. The electric field is solved by the Poisson equation. We 
have examined two cases of the background magnetic field,  0 and 
90 $\mu $G ($\Omega_{\rm ce}/\omega_{\rm pe}=0$ and 0.03), where 
$\Omega_{\rm ce}=eB/m_{\rm e} c$ is the electron cyclotron frequency. 
We sometimes refer the former and latter cases to the unmagnetized and 
magnetized cases, respectively. 
 
The size of the simulation box to the $x$- and $y$-directions is taken 
to be $L_{x}=64\lambda_{\rm Bun}$ {\rm and} $L_{y}=16\lambda_{\rm Bun}$, 
with a total of 2048 $\times$ 512 cells, where 
$\lambda_{\rm Bun}=2\pi \omega_{\rm pe}/v_{\rm d}$ is the wavelength 
of the most unstable mode of the Buneman instability.  
Thus, the length of each cell $\Delta x=\Delta y$ is 3 times the 
initial electron Debye length. The number of macroparticles is taken 
so that initially each cell includes 96 electrons and 96 total protons. 
The time step $\Delta t$ is taken as $5\times 10^{-3}\omega_{\rm pe}^{-1}$ 
and the simulation is followed until $3 \times10^3\omega_{\rm pe}^{-1}$ 
or $1.6 \times 10^{-3} \Omega_{\rm cp}^{-1}$ where $\Omega_{\rm cp}^{-1}$ 
corresponds to the time scale the upstream plasma stays in the foot region.

The differences from previous simulations \citep{ohi07} are as follows. 
First, the initial temperature is lower than the previous one $7{\rm eV}$. 
This is a more realistic one because the typical temperature of the 
interstellar matter is about 1eV. Secondly, we add returning proton beam 
in order that the total current vanishes, although in the electrostatic 
simulation it is not so critical. Thirdly, simulation time and simulation 
box are larger than the previous values so that we can investigate the 
ion two-stream instability. 

\subsection{Results}

Although we have performed simulations for two cases of the magnetic 
field strength ($0, 90\mu G$), the results turn out to be almost the 
same. Hence, we present the results of the unmagnetized case and add 
those of the magnetized case when necessary. 

First, we discuss the time development of the electric field. The 
evolution of the spatially averaged energy density of the electric 
field is shown in Figure 4. The solid and dashed curves show the 
$x$- and $y$-components, respectively, and bold and thin curves are 
unmagnetized and magnetized cases, respectively. In the first stage 
for $t<250\omega_{\rm pe}^{-1}$, the Buneman instability occurs and 
the electric field to both directions grow. After they attain peak 
values, they continue to decay till around 
$t\approx 10^3\omega_{\rm pe}^{-1}$. Then, after 
$t>10^3\omega_{\rm pe}^{-1}$, the $y$-component of the electric field 
starts to grow again while the $x$-component continues to decay. 
This feature is due to the ion two-stream instability as discussed below.

It should be noted that in the first stage of the Buneman instability 
($0<t<250\omega_{\rm pe}^{-1}$), the $y$-component of electric field 
is larger than the $x$-component. This is different from our previous 
result. As discussed in \S 2, this is because the temperature is lower 
and the waves with a larger obliqueness grow faster compared with our 
previous simulation \citep{ohi07}. In this stage, only electron 
temperature has risen up to about $m_{\rm e} v_{\rm d}^2$, but the 
ion temperature little changes (see Figure 5). Here, we define the 
temperature by the velocity dispersion, 
$T_{\rm s}\equiv m_{\rm s}\langle(v-\langle v\rangle)^2\rangle/2k_{\rm B}$.

In contrast to \citet{car88}, after the electrostatic waves caused by 
the Buneman instability decay at about $t=10^3 \omega_{\rm pe}^{-1}$, 
only the $y$-component of the electric field grows and oscillates 
after the amplitude saturates. In contrast, the $x$-component continues 
to decay. Of course, this feature can not be seen in one-dimensional 
simulations. The growth of the $y$-component of electric field is 
caused by highly oblique ion two-stream instability. If we consider 
only the parallelly propagating modes as in the one-dimesional simulations, 
ion acoustic instability has the largest growth rate. In the 
two-dimensional simulations we can take obliquely propagating modes 
with $k_{y}\neq 0$ into account, and the growth rate of the ion 
two-stream instability is larger than that of the ion acoustic 
instability as mentioned \S 2. A snapshot of the electrostatic 
potential at the saturation phase of the ion two-stream instabilities 
is shown in Figure 6. In Figure 6, spatial coordinates are normalized 
by $v_{\rm d}/\omega_{\rm pe}$, and the color contours show the 
electrostatic potential normalized by $m_{\rm e}v_{\rm d}^2/2$. 
Typical wave length scale to the $x$-direction is the order of the 
resonance scale $2\pi v_{\rm d} / \omega_{\rm pi}$ and that to 
the $y$-direction is comparable to $2\pi/k_{\rm D,e}$ as the linear 
analysis predicts. Before the saturation stage, the wavelength is 
smaller than $2\pi$ electron Debye length for which the linear growth 
rate is higher. 

Time development of the temperatures is shown in Figure 5. The 
electron temperature rises rapidly up to about $m_{\rm e} v_{\rm d}^2$ 
when the Buneman instability grows, while at this phase the proton 
temperature does not rise so much. At about 
$t=10^3 \omega_{\rm pe}^{-1}$, the proton temperature begins to rise 
by the growth of the ion two-stream instability but the electron 
temperature is kept at almost the same because this instability occurs 
between two ion beams and the energy density of the electric field is 
nearly 100 times smaller than the thermal energy density of electrons 
at this stage. At the end of the simulation, upstream proton 
temperature becomes about 100 times larger than the initial value 
and the electron to proton temperature ratio $T_{\rm e} / T_{\rm p}$ 
becomes about 10. For the magnetized case, $T_{\rm e} / T_{\rm p}$ 
becomes about 7. Although it is not explicitly shown in this paper, 
the proton distribution has a large anisotropy. Only the proton 
temperature of the $y$-direction rises up, but that of the 
$x$-direction is kept to be almost constant.
 
Figure 7 shows the energy distribution of electrons at the end of the 
simulation. The bold and thin curves represent non-mangetized and 
magnetized cases, respectively. It is noted that no high energy tail 
is seen. Even when there exists a magnetic field, electrons are not 
accelerated, that is, no surfing acceleration occurs in two-dimensional 
simulations. Both curves are of a flat top shape and can not be fitted 
by a Maxwellian distribution. If one draws Maxwellian distribution 
in Figure 7, it is a straight line with an inclination of 
$1/k_{\rm B}T_{\rm e}$. As discussed in \citet{ohi07}, at the end 
of simulation, the electron temperature becomes 
$T_{\rm e}\sim 0.5 m_{\rm e}v_{\rm d}^2$. These results are consistent 
with our previous result \citep{ohi07}. 

\section{Discussion}
Now we discuss the final outcome of the ion two-stream instability. 
It becomes unstable when two conditions are satisfied; one is the 
resonance condition, $k_{x} v_{\rm d} \sim \omega_{\rm pi}$ and the 
other is that the wave length be between the ion Debye length and 
the electron Debye length. Thus, the ion two-stream instability 
becomes stabilized when $T_{\rm p} \sim T_{\rm e}$. At the end of 
our simulation, the $y$-component of electric field has not decayed 
still completely. So, we expect $T_{\rm e}/T_{\rm p} < 10$ in the 
final stage, and probably ions will be heated up to 
$T_{\rm i} \sim m_{\rm e} v_{\rm d} ^2 \sim 4m_{\rm e} v_{\rm sh}^2$ 
at the foot region in high Mach number perpendicular shocks. 
Of course, because the proton temperature of the drift direction is 
still cold in the present simulation, we must check the isotropilazation 
process of ion velocity distribution by doing longtime full PIC-simulations.  

As for the electrons, the electron temperature in the foot region is 
also about $m_{\rm e} v_{\rm sh}^2$. In the later stage, the growth of 
the two-stream instability dominates over the ion acoustic instability 
and little electron heating occurs. Hence, as mentioned in \citet{ohi07}, 
if other electron heating mechanisms do not exist, after passing the 
shock front, electrons will undergo the adiabatic heating and finally, 
the electron temperature in the downstream becomes
\begin{equation}
T_{\rm e}\sim 4\times \frac{1}{2}m_{\rm e}(2v_{\rm sh})^2
= 0.41{\rm keV} \left(\frac{v_{\rm sh}}{0.01c}\right)^2, 
\end{equation}
where we assume that the compression ratio is 4. This has a very 
important implication for the electron heating process of the SNR shocks. 
The proton temperature in the downstream is 
$T_{\rm p}=3m_{\rm p}v_{\rm sh}^2/16$, hence the ratio of two temperatures 
is
\begin{equation}
T_{\rm e}/T_{\rm p} \sim \frac{128}{3}\frac{m_{\rm e}}{m_{\rm p}} \sim 0.023. 
\end{equation}
This value is close to the observed value as long as the shock velocity 
$v_{\rm sh}$ is larger than $1500{\rm km/s}$ \citep{ade08}. Namely, 
we expect that the overheating problem of electrons raised by 
\citet{car88} can be solved by the ion two-stream instability.

In this paper, we prepare three ion beams as the initial condition. We 
have also performed simulations for other initial conditions such that 
there exist upstream electrons, upstream and reflected protons. Electrons 
and reflected protons have drift velocities to satisfy the vanishing 
current condition. This situation corresponds to the initial phase of the 
shock reformation phenomena. 
The results turn out to be basically the same as those in the results 
presented in this paper.

Our simulation does not include electromagnetic modes. In the linear 
stage of the Buneman and ion two-tream instabilities, the effects are 
negligible because the growth rates of  electromagnetic modes are 
smaller than those of electrostatic modes for $v_{\rm d} \sim 0.01c$. 
In the nonlinear stage, electromagnetic modes may be important. One of 
the reasons is that the electrostatic waves with a wave vector almost 
perpendicular to the drift direction may make currents because the 
charge fluctuation can not be shielded by electrons in this situation, 
where the fluctuation scale is smaller than the electron Debye length. 
Consequently the current may make the magnetic field. It is an interesting 
speculation that the magnetic field might be amplified more rapidly by 
the ion two-stream instability than the ion Weibel instability. The 
other reason is that anisotropic ion heating caused by the highly 
oblique ion two-stream instability excites the Weibel instability due 
to ion temperature anisotropy. At the end of our simulations, the ion 
temperature of $x$-direction is almost the same as initial one and the 
ratio of the ion temperature of $y$-direction to that of $x$-diretion 
$T_{\rm iy}/T_{\rm ix}$ is about 100. These two features may lead to 
magnetic field amplification and accompanying particle acceleration and 
heating in the shock foot region. We will make full-PIC simulations in 
future work to investigate these issues.
 
\section{Summary}
We performed linear analysis of two-dimensional electrostatic modes 
and electrostatic two-dimensional PIC simulations with the real proton 
electron mass ratio to investigate the time evolution of electrostatic 
waves in the foot region of collisionless shocks with a high Mach number. 
We consider only the foot region by properly modeling the effects of 
reflected and returning protons. Performing the linear analysis, we 
have shown that after electrons are heated by the Buneman instability, 
the fastest growing mode is not the ion acoustic instability but the 
highly oblique ion two-stream instability. The latter mode, which has 
not been noticed previously, is excited by the resonace between 
ion plasma oscillations of the two proton beams when the electron 
temperature is much higher than the ion temperature. The PIC simulation 
confirms that the excitaion of the ion two-stream instability occurs 
faster than the ion acoustic instability and that protons are heated 
preferentially to the perpendicular direction to the shock normal 
direction. As a result, electron heating basically stops at the stage of 
the Buneman instability and the expected electron temperature in 
supernova remnants is fully compatible with the observation, avoiding 
the overheating problem raised by \citet{car88}. 

\acknowledgments
We are grateful to T. Tsuribe, T. Umeda, R. Yamazaki, T. Kato, 
M. Hoshino, T. Terasawa, S. Matsukiyo for discussions and suggestions 
especially for providing useful guidance and information in doing 
numerical simulations and for the nature of the ion two-stream instability. 
This work is partly supported by Scientific Research Grants (F.T.: 18542390
and 20540231) by the Ministry of Education, Culture, Sports, Science and Technology 
of Japan. Y. O. is supported by a Grant-in-Aid for JSPS Research Fellowships for Young Scientists. Numerical computations were carried out on fpc-cluster system at Center for Computational Astrophysics, CfCA, of National Astronomical Observatory of Japan.

\clearpage

\begin{figure}
\plottwo{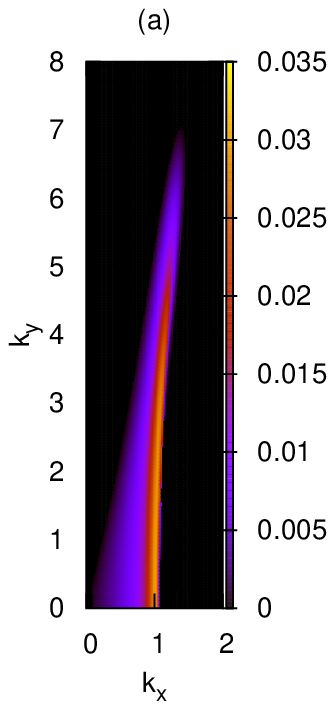}{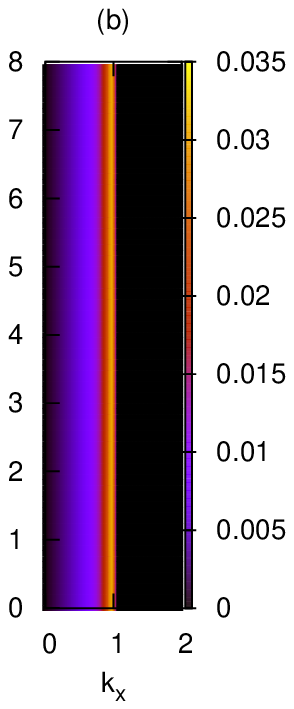}
\caption{The color contour plot of the growth rate of the Buneman instabilitiy. The left panel (a) is for $v_{\rm d}/v_{\rm th,e}=10, T_{\rm e} = T_{\rm p}$, while the right panel (b) is for the  cold limit. \label{fig1}}
\end{figure}


\begin{figure}
\plottwo{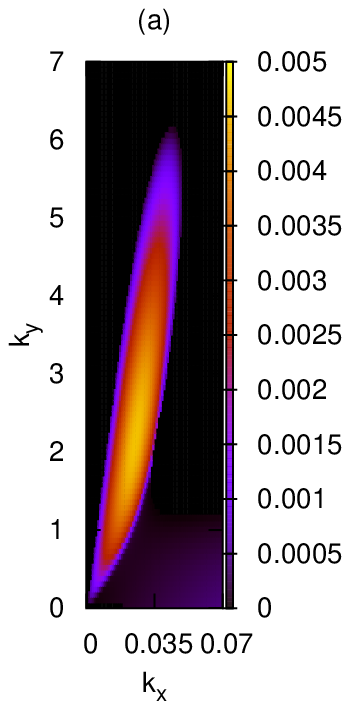}{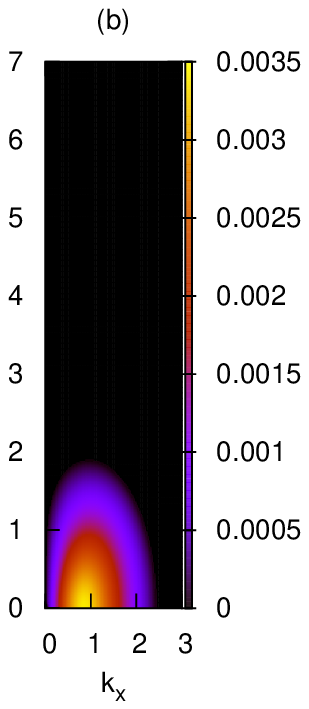} 
\caption{The color contour plot of the growth rate of the ion two-stream instability (the left panel (a)) and that of the ion acoustic instability, for $v_{\rm d}/v_{\rm th,e}=1, T_{\rm e} = 100T_{\rm p}$. Note the diffenence in the scale of the abscissa between (a) and (b).
\label{fig2}}
\end{figure}


\begin{figure}
\plotone{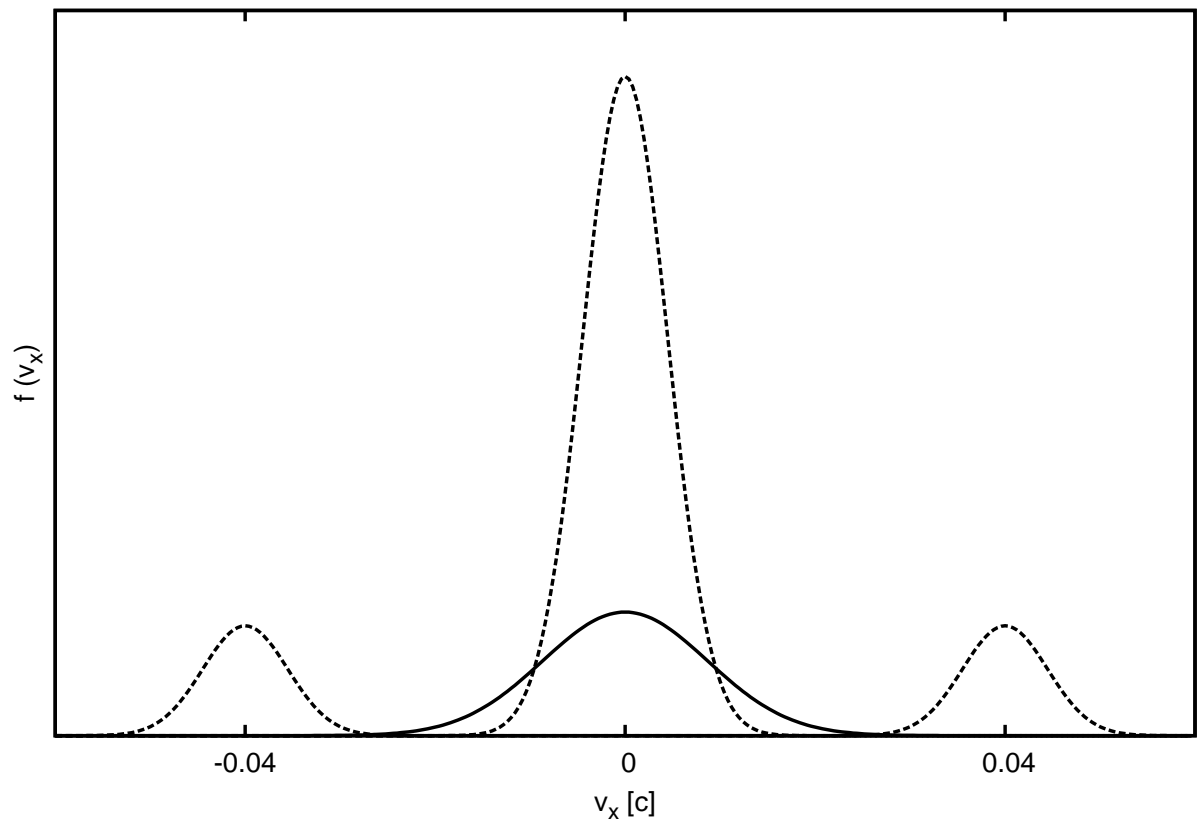} \caption{The particle distribution in the initial state for the simulation. Solid and dashed curves represent electron and ion distribution functions of $v_x$, respectively. 
\label{fig3}}
\end{figure}


\begin{figure}
\plotone{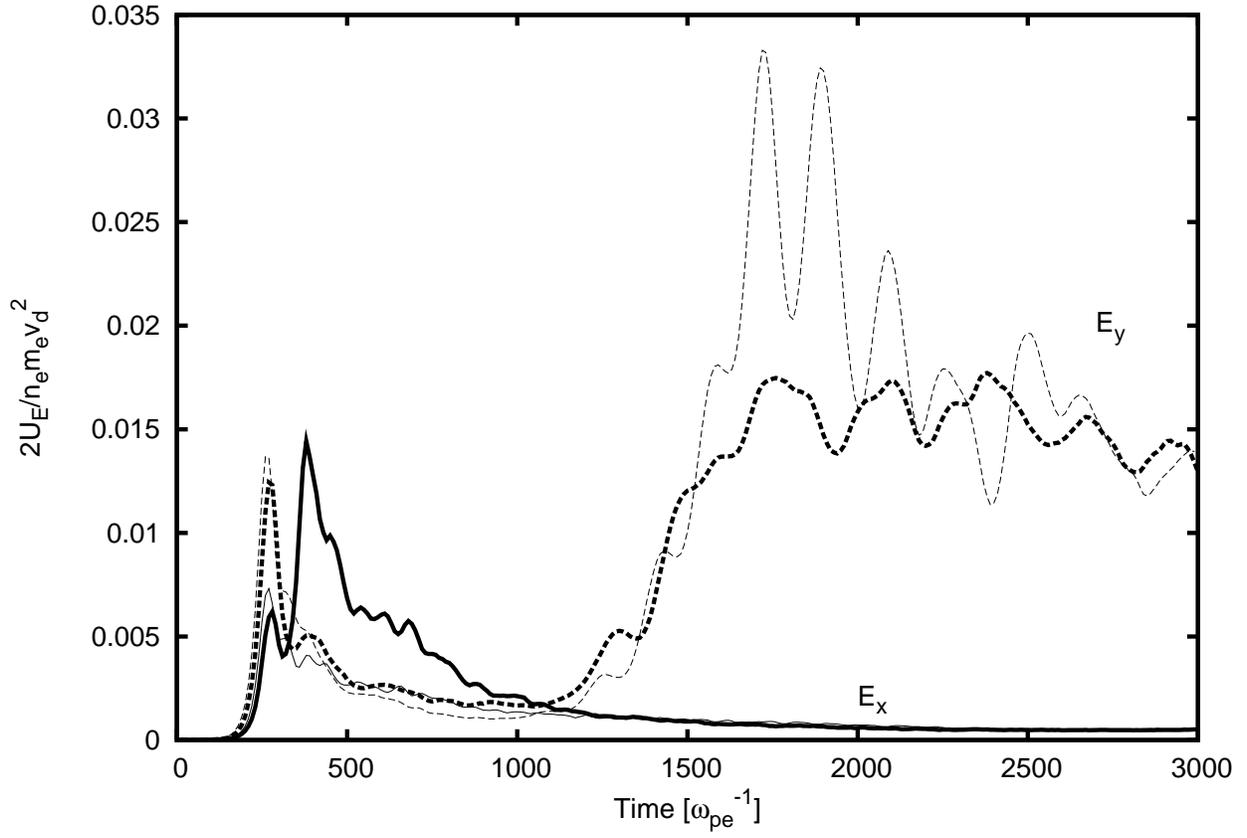} 
\caption{The development of the energy density of electric field. 
The bold and thin lines represent non-magnetized and magnetized cases, respectively. Solid and dashed curves represent $E_x$ and $E_y$, respectively. 
\label{fig4}}
\end{figure}


\begin{figure}
\plotone{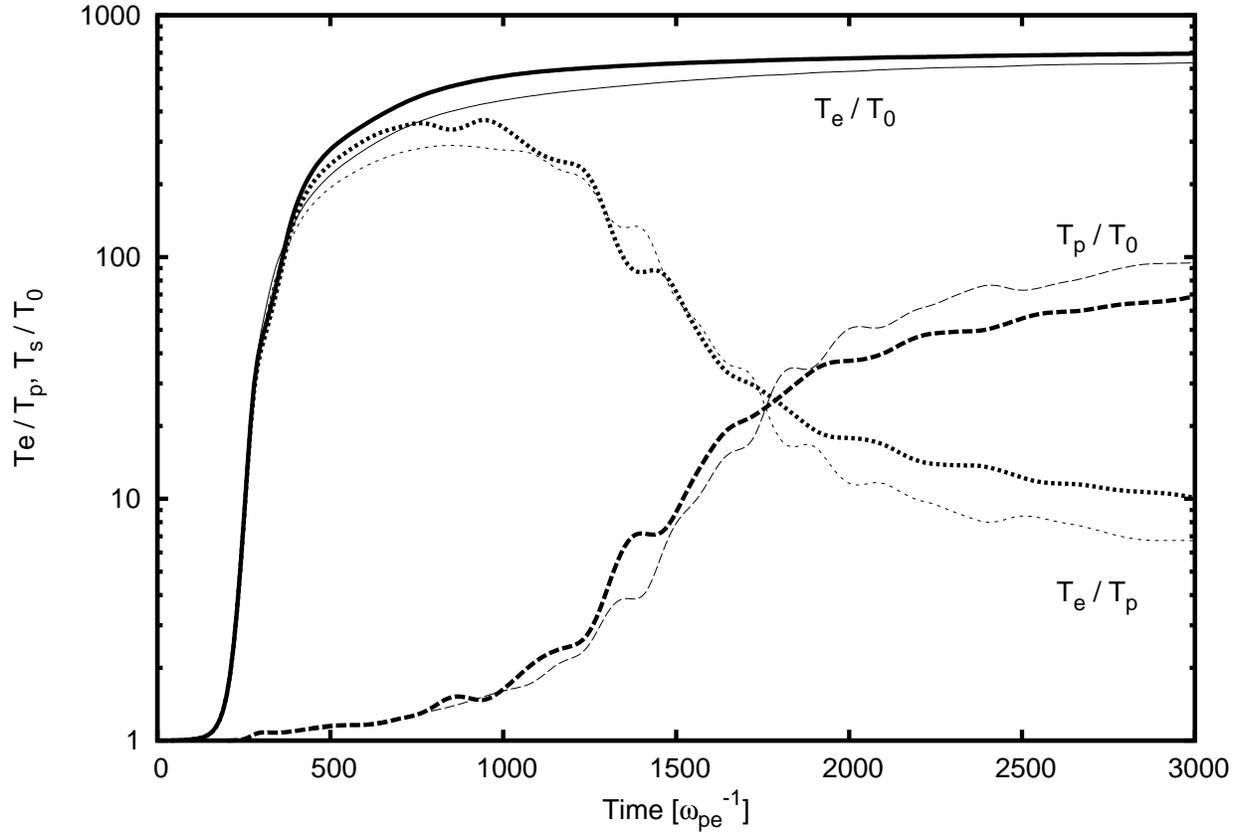} 
\caption{Time development of the temperatures. 
The bold and thin lines represent non-magnetized and magnetized cases, respectively. Solid and dashed curves represent electron and proton temperatures, respectively. Dotted curves represent electron proton temperature ratio. 
\label{fig5}}
\end{figure}


\begin{figure}
\plotone{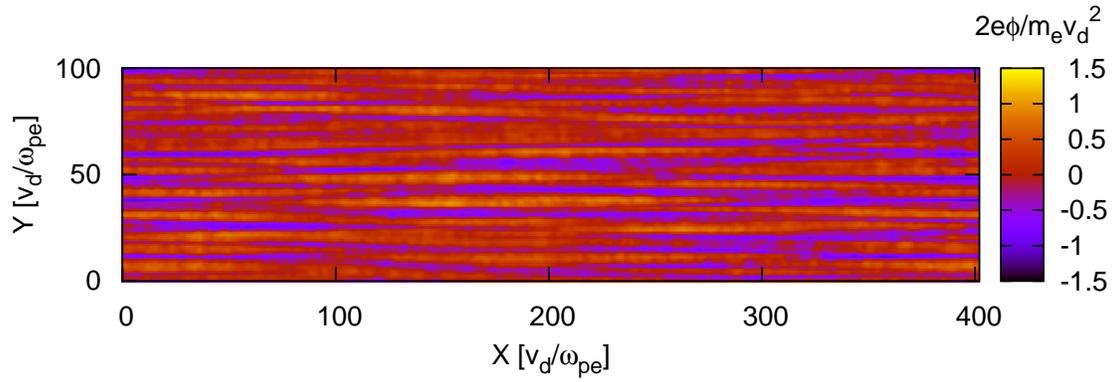} 
\caption{The contour of the electrostatic potential at $t=1740 \omega_{\rm pe}^{-1}$.The potential is normalized by $m_{\rm e}v_{\rm d}^2/2$. Note that the structure is filamentary so that the electric field is almost perpendicular to the shock normal direction.
\label{fig6}}
\end{figure}


\begin{figure}
\plotone{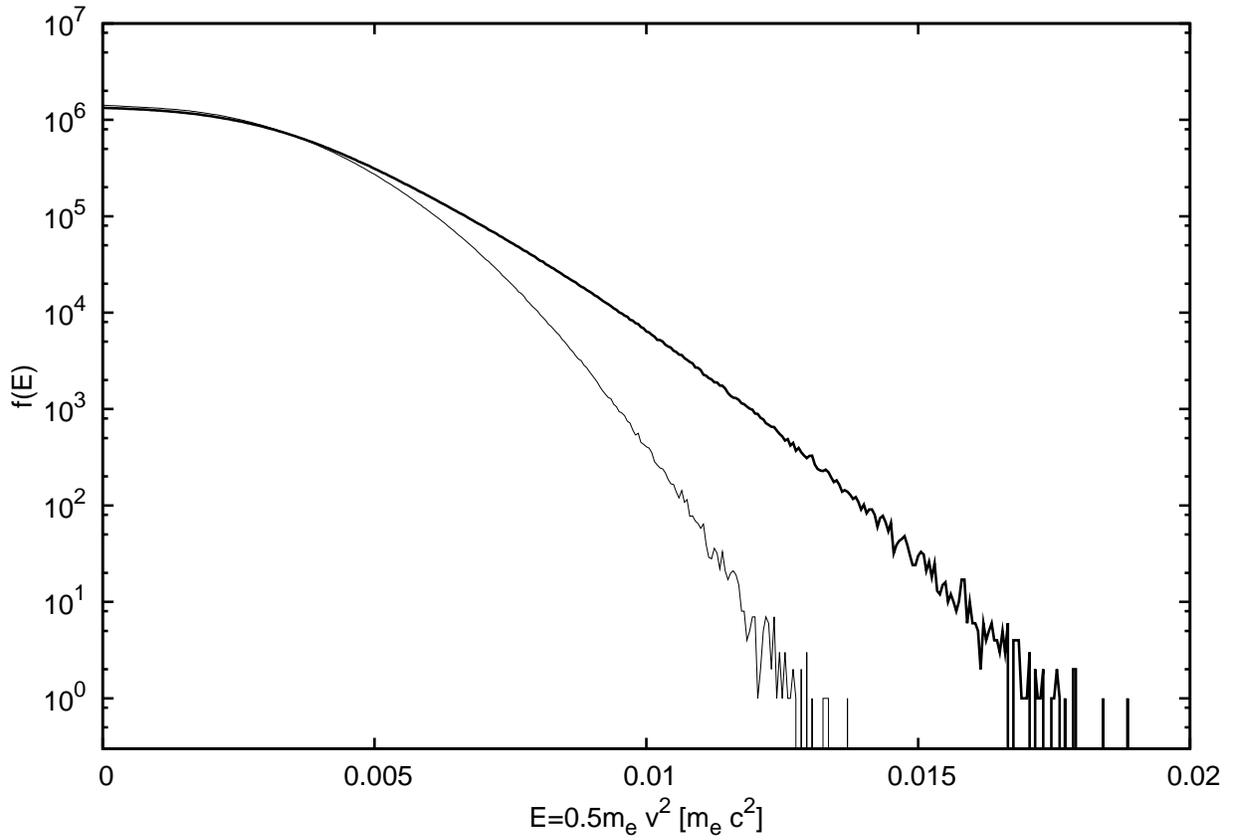} 
\caption{Energy distribution function of electrons at the end of simulations. Bold and thin curves represent non-magnetized and magnetized cases, respectively. 
\label{fig7}}
\end{figure}

\end{document}